\documentclass[aps,prl,twocolumn,reprint,amsmath,amssymb,superscriptaddress,floatfix,footinbib,longbibliography]{revtex4-2}

\usepackage{graphicx}
\usepackage{epstopdf}
\usepackage{rotating}
\usepackage[T1]{fontenc}
\usepackage[applemac]{inputenc}
\usepackage{lmodern}
\usepackage[english]{babel}
\usepackage{booktabs}
\usepackage{ae}
\usepackage{siunitx}

\usepackage{amsmath,amssymb,natbib,bm}
\usepackage{psfrag}
\usepackage{amsthm}
\usepackage{algorithm}
\usepackage{algpseudocode}

\usepackage{xcolor, colortbl}
\usepackage{multirow}
\usepackage[colorlinks]{hyperref}
\hypersetup{%
        plainpages=true,
        breaklinks=true,
        hypertexnames=false,
        pageanchor=true,
        colorlinks=true,
        linkcolor={blue},
        citecolor={blue},
        urlcolor={blue},
        anchorcolor={black}
      }

\usepackage{mleftright} 
\usepackage{algorithm}
\usepackage{algpseudocode}

\newcommand{\figref}[1]{\mbox{Fig.~\ref{#1}}}

\renewcommand{\eqref}[1]{\mbox{Eq.~(\ref{#1})}}

\newcommand{\figpanel}[2]{Fig.~\hyperref[#1]{\ref*{#1}(#2)}}
\newcommand{\figpanels}[3]{Fig.~\hyperref[#1]{\ref*{#1}(#2)--(#3)}}
\newcommand{\figpanelNoPrefix}[2]{\hyperref[#1]{\ref*{#1}(#2)}}

\newcommand{\ket}[1]{|#1\rangle}

\newcommand{\be}{\begin{equation}}
\newcommand{\ee}{\end{equation}}
\newcommand{\bea}{\begin{eqnarray}}
\newcommand{\eea}{\end{eqnarray}}

\usepackage{color}

\usepackage{soul}
\usepackage{verbatim}

\begin{document}

\author{Aniket Patel}
\affiliation{Department of Microtechnology and Nanoscience, Chalmers University of Technology, 41296 Gothenburg, Sweden}

\author{Akshay Gaikwad}
\affiliation{Department of Microtechnology and Nanoscience, Chalmers University of Technology, 41296 Gothenburg, Sweden}

\author{Tangyou Huang}
\affiliation{Department of Microtechnology and Nanoscience, Chalmers University of Technology, 41296 Gothenburg, Sweden}

\author{Anton Frisk Kockum}
\affiliation{Department of Microtechnology and Nanoscience, Chalmers University of Technology, 41296 Gothenburg, Sweden}

\author{Tahereh Abad}
\email{tahereh.abad@chalmers.se}
\affiliation{Department of Microtechnology and Nanoscience, Chalmers University of Technology, 41296 Gothenburg, Sweden}

\title{Selective and efficient quantum state tomography for multi-qubit systems}

\begin{abstract}
Quantum state tomography (QST) is a crucial tool for characterizing quantum states. However, QST becomes impractical for reconstructing multi-qubit density matrices since data sets and computational costs grow exponentially with qubit number. In this Letter, we introduce selective and efficient QST (SEEQST), an approach for efficiently estimating multiple selected elements of an arbitrary $N$-qubit density matrix. We show that any $N$-qubit density matrix can be partitioned into $2^N$ subsets, each containing $2^N$ elements. With SEEQST, any such subset can be accurately estimated from just two experiments with only single-qubit measurements. The complexity for estimating any subset remains constant regardless of Hilbert-space dimension, so SEEQST can find the full density matrix using $2^{N+1} - 1$ experiments, where standard methods would use $3^N$ experiments. We provide a circuit decomposition for the SEEQST experiments, demonstrating that their maximum circuit depth scales logarithmically with $N$ assuming all-to-all connectivity. The Python code for SEEQST is publicly available at   
\href{https://github.com/aniket-ae/SEEQST}{github.com/aniket-ae/SEEQST}.

\end{abstract}

\date{\today}

\maketitle

\paragraph*{Introduction.}

Quantum state tomography (QST) reconstructs the complete density matrix of a quantum system by measuring the expectation values of observables and performing computational post-processing~\cite{james-pra-2001, Paris2004, liu-prb-2005, lvovsky-rmp-2009, cramer-prl-2013}. Such characterization of quantum states is crucial for the development of quantum technologies~\cite{Gebhart2023, Hashim2024, Blume-Kohout2025}. However, despite its conceptual simplicity, QST becomes impractical for large systems, since data and computational demands increase rapidly with system size~\cite{riofrio-natcom-2017, li-pra-2017, cotler-prl-2020}. 

Well-established and widely used approaches for QST include maximum likelihood estimation~\cite{ban-pra-1999, imle, mle-pra-2007, smolin-prl-2012, shang-pra-2017}, gradient-descent algorithms~\cite{ferrie-prl-2014, bolduc-npj-2017, hsu-prl-2024, wang-prr-2024, Gaikwad2025, Hoshi2025}, machine learning and deep neural networks~\cite{quek-npj-2021, lohani-mlst-2020, gaikwad-pra-2024, schmale-2022, Ahmed-cgan-2021, Ahmed2021a, yu-2019, innan-2024}, variational algorithms~\cite{xin-2020, liu-pra-2020}, convex optimization techniques~\cite{david-prl-2010, Steffens_2017, ingrid-prapp-2022, gaikwad-qip-2021, nehra-prr-2020}, and many others~\cite{cramer-natcom-2010, lanyon-np-2017, toth-prl-2010, Kyrillidis2018}. While these methods try to speed up QST, primarily by focusing on the post-processing step (reconstructing the full-density matrix from collected data), they all suffer from the fundamental limitations of growing requirements for data and computation. However, there remains significant room for devising flexible and information-theoretic approaches to the data acquisition itself~\cite{flammia-prl-20111, huang-natphys-2020, anushya-prl-2024, bonet-prx-2020}. Such strategic data collection could enable efficient partial state reconstruction, offering practical benefits and insights into quantum systems without full QST. This is the approach we develop in this Letter.

Recent approaches to estimating selected elements of density matrices have primarily relied on weak measurements (WMs)~\cite{lundeen-prl-2012, wu-sr-2013, gaikwad-epjd-2023, kim-natcom-2017, luca-prl-2018, lundeen-prl-2016, zhang-prl-2019}. While WMs enable direct extraction of specific density-matrix elements without complex data post-processing, they require pure ancilla qubits and complex multi-qubit entangling operations, making them resource-intensive and challenging to implement~\cite{gaikwad-epjd-2023}. Additionally, WM strategies are inherently approximate and suffer from a trade-off: validity of the approximation necessitates weaker interactions, while reducing statistical uncertainties requires stronger ones.

Beyond WM schemes, few alternative protocols have been developed~\cite{feng-pra-2021, li-pra-2022, ekert-prl-2002, paz-pra-2013, Bolduc-natcom-2016, Bendersky-pra-2022, morris-arxiv-2020, Feldman2024, Laurell2025}. Phase-shifting schemes~\cite{feng-pra-2021, li-pra-2022} can estimate a single density-matrix element of a multi-qubit system using only six expectation values. In Ref.~\cite{ekert-prl-2002}, a fixed controlled-SWAP network was used to compute density-matrix functionals, requiring ancilla qubits. Selective QST has also been explored for both discrete- and continuous-variable systems, where precision depends on the number of experiments~\cite{paz-pra-2013, Bendersky-pra-2022, kevin-natcom-2024, Feldman2024}. Additionally, direct measurement techniques have been used to reconstruct quantum state coefficients of a $10^5$-dimensional entangled state in an optical setup~\cite{Bolduc-natcom-2016}. A recently reported partial quantum shadow tomography scheme \cite{mahesh-arxiv-2025} enables the estimation of a subset of density matrix elements.  Other works explore experimental requirements for pure-state tomography; however, they inherently lack selectivity~\cite{goyeneche-prl-2015, feng-arxiv-2025}. Despite these advancements, a universal, scalable approach for selectively estimating density-matrix elements remains elusive. Existing protocols are often constrained by system size, accuracy, and precision, scaling unfavorably with Hilbert-space dimension.

In this Letter, we address this fundamental challenge by introducing selective and efficient quantum state tomography (SEEQST). Using only two carefully designed experiments, SEEQST enables efficient estimation of specific density-matrix elements. We demonstrate that an $N$-qubit density matrix can be systematically partitioned into $2^N$ subsets, each containing a predefined set of $2^N$ elements, and that any chosen subset can be accurately estimated using our SEEQST protocol. Notably, SEEQST maintains constant complexity per subset, independent of Hilbert-space dimension, making it highly scalable for multi-qubit systems. We provide a circuit decomposition for SEEQST, showing that its maximum circuit depth scales logarithmically with the number of qubits for all-to-all connectivity. Unlike traditional QST methods, which require processing the entire data set at once to reconstruct the quantum state, SEEQST's independent subset computation allows parallel estimation for efficient full QST. This significantly reduces computational overhead and enhances scalability, establishing SEEQST as a powerful alternative to conventional QST techniques.

We provide a Python code to implement our SEEQST~\cite{SEEQST-python-code} protocol. The first part of the code takes a list of density-matrix elements as input and outputs the required experiments. The second part uses the data from these proposed experiments as input to estimate the elements. Our code also includes a version of SEEQST based on only local operations.

\paragraph*{Method.}

An $N$-qubit density matrix $\rho$ has elements $\rho_{ij} = \langle i | \rho | j \rangle$, where $\{ |i\rangle \}$ are the computational basis vectors; $i \in [0, 1, \dots, 2^N-1] $, such that $\rho_{ij}$ corresponds to the element in the $(i+1)$th row and $(j+1)$th column. To compute $\rho_{ij}$, one must evaluate the expectation value of the operator $ \Pi_{ij} = |j\rangle \langle i|$. In general, $ \Pi_{ij}$ is a non-Hermitian operator, except when $i = j$, which yields diagonal elements. This underscores the challenge of selectively estimating off-diagonal elements of $\rho$. A straightforward approach is to express $\Pi_{ij}$ as $\sum_m a_m E_m$, where $a_m$ are complex coefficients and $\{ E_m \}$ are Hermitian operators forming a basis, typically the $N$-qubit Pauli set $E_m = \{ I, \sigma_x, \sigma_y, \sigma_z \}^{\otimes N}$. Measuring $\langle E_ m \rangle$ for nonzero $a_m$ then allows reconstruction of $\langle \Pi_{ij} \rangle$ via $\sum_m a_m \langle E_m \rangle $. This approach requires fewer experiments than full QST, but the number still scales exponentially with the number of qubits, due to the lack of a well-defined link between target elements and the required measurements~\cite{gaikwad-pra-2018, gaikwad-sr-2022}. 

With SEEQST, we provide a more efficient strategy to collectively compute these operators instead of measuring them individually, minimizing the number of required experiments. We first illustrate how this works for single- and two-qubit systems, before generalizing to $N$ qubits.

In single-qubit SEEQST, the $2 \times 2$ density matrix is partitioned into two sets:  
$S_1 = \{ \rho_{00}, \rho_{11} \}$ (diagonal elements, $i = j$) and $S_2 = \{ \rho_{01}, \rho_{10} \}$ (off-diagonal elements, $i \neq j$). The elements of $S_1$ are determined by measuring $\{ I, \sigma_z \}$; those in $S_2$ are estimated using $\{ \sigma_x, \sigma_y \}$. Here, the identity operator $I$ is redundant, as its expectation value is always $1$, so $S_1$ can be determined in a single experiment by just measuring $\sigma_z$. In contrast, estimating $S_2$ requires two separate experiments, one for measuring $\sigma_x$ and another for $\sigma_y$, since $[\sigma_x, \sigma_y] \neq 0$. Thus, for a single qubit, SEEQST is essentially equivalent to standard full QST. However, this equivalence does not hold for multi-qubit systems, as demonstrated next. 

\begin{figure} 
\centering 
\includegraphics[width= \linewidth]{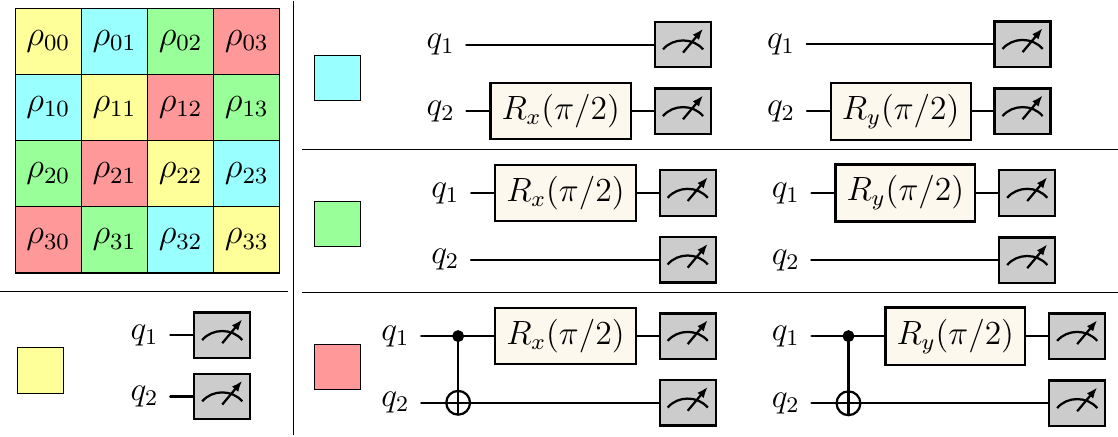} 
\caption{Classification of two-qubit density-matrix elements based on the parity-preserving structure of indices required for SEEQST. Each class is represented by a color (yellow, blue, green, or red) and two corresponding quantum circuits for measurements determining all density-matrix elements in the class (one experiment for the diagonal elements).}
\label{fig:2qubit} 
\end{figure}

For two-qubit SEEQST, the density-matrix elements are $\rho_{ij} = \rho_{i_1 i_2, j_1 j_2} = \langle i_1 i_2 | \rho | j_1 j_2 \rangle$, where $i_1 i_2$ and $j_1 j_2$ are binary representations of $i$ and $j$, respectively, labeling the two-qubit computational basis states. The $4 \times 4$ density matrix can be partitioned into four sets with four elements each, $\{ S_1, S_2, S_3, S_4\}$, represented by different colors in \figref{fig:2qubit}, based on whether the indices corresponding to each qubit, i.e., $i_1$ and $j_1$ (or $i_2$ and $j_2$), are identical or not. Specifically, $S_1 = \{ \rho_{i_1 i_2, j_1 j_2} \vert i_1 = j_1 , i_2 = j_2 \}$ consists of all diagonal elements (yellow in \figref{fig:2qubit}), while $S_2 = \{ \rho_{i_1 i_2, j_1 j_2} \vert i_1 = j_1 , i_2 \neq j_2 \}$ has the elements that are diagonal in the first qubit but off-diagonal in the second (blue in \figref{fig:2qubit}). Similarly, $S_3 = \{ \rho_{i_1 i_2, j_1 j_2} \vert i_1 \neq j_1 , i_2 = j_2 \}$ and $S_4 = \{ \rho_{i_1 i_2, j_1 j_2} \vert i_1 \neq j_1 , i_2 \neq j_2 \}$ (green and red, respectively, in \figref{fig:2qubit}). The elements in any set can thus be obtained from a single element by preserving the parities of the pairs of indices $(i_1,j_1)$ and $(i_2,j_2)$ through the transformation $\{ I, X\}^{\otimes 2} (i_1 i_2, j_1 j_2) \rightarrow (i'_1 i'_2, j'_1 j'_2)$, where $X$ denotes a bit-flip gate and the first (second) operator acts on both $i_1$ ($i_2$) and $j_1$ ($j_2$). 

This partitioning of the density-matrix elements into sets provides a systematic framework for reconstructing $\rho$ and selectively estimating desired elements $\rho_{i_1 i_2,j_1 j_2}$. The partitioning reflects the decomposition of $\rho$ in the Pauli operator basis: diagonal elements $(i_{1(2)} = j_{1(2)})$ correspond to $\{I, \sigma_z\}$, while off-diagonal elements $(i_{1(2)} \neq j_{1(2)})$ are associated with $\{\sigma_x, \sigma_y\}$. Hence, the optimal observables for each set are $\Pi^{S_1} = \{I, \sigma_z\} \otimes \{I, \sigma_z\}$, $\Pi^{S_2} = \{I, \sigma_z\} \otimes \{\sigma_x, \sigma_y\} $, $\Pi^{S_3} = \{\sigma_x, \sigma_y\} \otimes \{I, \sigma_z\}$, and $\Pi^{S_4} = \{\sigma_x, \sigma_y\} \otimes \{\sigma_x, \sigma_y\}$. Together, these sets span all sixteen Pauli matrices, forming a tomographically complete set for full QST.

We now demonstrate that just two measurement settings are sufficient to determine any entire set $S_k$. For instance, $\Pi^{S_3} = \{\sigma_x \otimes I , \sigma_x \otimes \sigma_z, \sigma_y \otimes I, \sigma_y \otimes \sigma_z \}$ can be divided into two subsets of mutually commuting observables: $\mathcal{E}(\Pi^{S_3}) = \{ \sigma_x \otimes I , \sigma_x \otimes \sigma_z  \}$ and $\mathcal{O}(\Pi^{S_3}) = \{\sigma_y \otimes I, \sigma_y \otimes \sigma_z\}$. Each subset corresponds to a distinct measurement setting in the common eigenbasis of $\mathcal{E}(\Pi^{S_3})$ and $\mathcal{O}(\Pi^{S_3})$. The required quantum circuits are easily constructed by implementing basis transformations associated with $\mathcal{E}(\Pi^{S_3})$ and $\mathcal{O}(\Pi^{S_3})$, utilizing their normalized common eigenstates,
\begin{align}
\mathcal{E}(\Pi^{S_3}): \quad |e^{\pm}_q\rangle &= \frac{1}{\sqrt{2}} (|0\rangle \pm |1\rangle)|q\rangle , \label{eq:e_eigenstate_2q} \\
\mathcal{O}(\Pi^{S_3}):\quad |o^{\pm}_q\rangle &= \frac{1}{\sqrt{2}} (|0\rangle \pm i |1\rangle)|q\rangle , \label{eq:o_eigenstate_2q}
\end{align}
where $q\in \{|0\rangle, |1\rangle\}$, followed by standard measurements in the computational basis. The basis transformations for $\mathcal{E}(\Pi^{S_3})$ and $\mathcal{O}(\Pi^{S_3})$ are $U_{\mathcal{E}} = \begin{bmatrix}
\ket{e^{+}_0} & \ket{e^{-}_0} & \ket{e^{+}_1} & \ket{e^{-}_1}
\end{bmatrix}^{\dagger} $  and $U_{\mathcal{O}} = \begin{bmatrix}
\ket{o^{+}_0} & \ket{o^{-}_0} & \ket{o^{+}_1} & \ket{o^{-}_1}
\end{bmatrix}^{\dagger}$. Alternative transformations can be obtained by permuting columns within $U_{\mathcal{E}}$ and $U_{\mathcal{O}}$. However, in this particular case, the unitary operators decompose into local operations,  $U_{\mathcal{E}} = R_y(\pi/2) \otimes I$ and $U_{\mathcal{O}} = R_x(\pi/2) \otimes I$, simplifying implementation. A similar approach applies to $S_1$, $S_2$, and $S_4$. For the diagonal elements ($S_1$), only a single experiment is needed, as all observables in the corresponding set mutually commute. For $S_4$, both experiments involve a CNOT gate, resulting in the maximum circuit depth.

\paragraph*{Main result.}

In $N$-qubit SEEQST, the density matrix elements are $\rho_{ij} = \langle i_1 i_2 \cdots i_N | \rho | j_1 j_2 \cdots j_N \rangle$, where $i_1 i_2 \cdots i_N$ and $j_1 j_2 \cdots j_N$ are the binary expansions of $i$ and $j$, respectively. The $2^N \times 2^N$ density matrix can be systematically partitioned into $2^N$ sets with $2^N$ elements each, $\{ S_1, S_2, \dots, S_{2^N} \}$, based on whether $i_l = j_l$ or $i_l \neq j_l$ for each qubit index $l$. Each set $S_k$ is spanned by any single element within it through the transformations $\{ I, X\}^{\otimes N}(i_1 i_2 \cdots i_N, j_1 j_2 \cdots j_N) \rightarrow (i'_1 i'_2 \cdots i'_N, j'_1 j'_2 \cdots j'_N)$ that preserve the parities for each pair of indices $(i_l, j_l)$. 

For a given set $S_k$, the optimal set of $N$-qubit Pauli observables that contribute non-trivially is
\begin{equation}\label{eq:sqstset}
   \Pi^{S_k}  = \bigotimes_{l=1}^N 
    \begin{cases}
        \{I, \sigma_z\}, & \text{if } (i_l, j_l) \in \{(0,0), (1,1)\}, \\
        \{\sigma_x, \sigma_y\}, & \text{if } (i_l, j_l) \in \{(0,1), (1,0)\}.
    \end{cases}
\end{equation}
In the trivial case of diagonal elements, $S_1 = \{ \rho_{i, j} \vert i_l=j_l \:\forall l \}$, the associated set of observables, $\Pi^{S_1} = \{I, \sigma_z\}^{\otimes N}$, consists of mutually commuting operators. Consequently, all elements in $S_1$ can be determined in a single experiment: a standard measurement in the computational basis. On the other hand, the set of observables $\Pi^{S_k}$ for an $S_k$ with some off-diagonal elements, where $M$ qubits satisfy $i_l \neq j_l$ and the remaining $N-M$ qubits have $i_l =j_l$, can be divided into two subsets of mutually commuting observables, $\mathcal{E}(\Pi^{S_k})$ and $\mathcal{O}(\Pi^{S_k})$. Specifically, $\mathcal{E}(\Pi^{S_k})$ contains Pauli tensor products with an even number of $\sigma_y$, while $\mathcal{O}(\Pi^{S_k})$ includes those with an odd number of $\sigma_y$. This division ensures that all observables within each set can be measured in a single experiment. We provide a formal proof of this statement in the Supplementary Material (SM)~\cite{SuppMat}. 

Relabeling the $M$ qubits with $i_l \neq j_l$ as the first qubits and the other $N-M$ qubits as the last, the simultaneous eigenstates of $\mathcal{E}(\Pi^{S_k})$ and $\mathcal{O}(\Pi^{S_k})$ can be expressed as  
\begin{align}
 \mathcal{E}(\Pi^{S_k}): \quad   |e_{p\overline{p}q}^{\pm}\rangle &= \frac{1}{\sqrt{2}} (|p\rangle \pm |\overline{p}\rangle) |q\rangle, \label{eq:e_eigenstate} \\
 \mathcal{O}(\Pi^{S_k}): \quad   |o_{p\overline{p}q}^{\pm} \rangle &= \frac{1}{\sqrt{2}} (|p\rangle \pm i |\overline{p}\rangle) |q\rangle, \label{eq:o_eigenstate}
\end{align}
where $|p\rangle \in |0\rangle \otimes \{|0\rangle, |1\rangle\}^{\otimes (M-1)}$, $|\overline{p}\rangle$ is its bit-wise complement, and $|q\rangle \in \{|0\rangle, |1\rangle\}^{\otimes (N-M)}$. Although $|p\rangle \in \{ |0\rangle, |1\rangle \}^{\otimes M}$ also is a valid set of eigenstates, this choice leads to redundancy; to avoid repeated eigenstates, we must fix one of the qubits to either $|0\rangle $ or $|1\rangle$. 

\begin{figure} 
\centering 
\includegraphics[width=0.8 \linewidth]{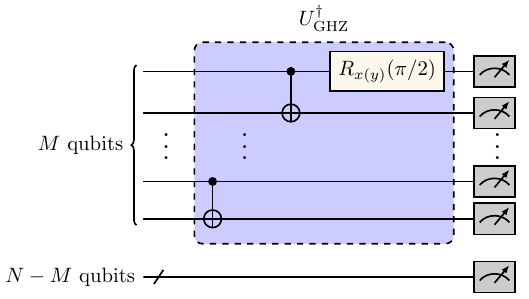} 
\caption{Quantum circuits to implement $N$-qubit SEEQST. Choosing $R_x(\pi/2)$ or $R_y(\pi/2)$ in the depicted circuit implements the basis transformation $U_{\mathcal{O}}$ or $U_{\mathcal{E}}$, respectively. The first $M$ qubits having $i_l \neq j_l$ (off-diagonal case) are acted upon by $U^{\dagger}_{\rm GHZ}$ (light purple shading). The remaining $N-M$ qubits, represented by a multi-qubit quantum register with '/', correspond to $i_l = j_l$ (diagonal case).
} 
    \label{fig:nqubit} 
\end{figure}

The basis transformations $U_{\mathcal{E}}$ and $U_{\mathcal{O}}$ map the computational basis to Greenberger--Horne--Zeilinger (GHZ)-type~\cite{Greenberger1989, Dur2000} eigenstates as shown in Eqs.~(\ref{eq:e_eigenstate})--(\ref{eq:o_eigenstate}). These transformations are thus the conjugate transpose of the unitary operations $U_{\rm GHZ}$ used to prepare $M$-qubit GHZ states. The potential advantage of measurements in a GHZ-type entangled basis has been demonstrated in the context of reducing the sample complexity for learning Pauli expectation values~\cite{matteo-quantum-2024}.
These operations consist of a local rotation $R_{x(y)}(\pi/2)$ on one qubit, followed by $M-1$ CNOT gates, as illustrated in Fig.~\ref{fig:nqubit}, giving a maximum circuit depth that grows linearly with the number of qubits. However, the circuit depth can be reduced to $O(\log M)$ for all-to-all qubit connectivity by parallelizing the CNOT gates~\cite{KamGHZdepth, mooney-jpc-2021, feng-prapplied-2022, liao-arxiv-2024}, as explained in the SM~\cite{SuppMat}. Without entangling gates, SEEQST reduces to a constrained version of standard QST, requiring $3^N$ circuits for estimating the full density matrix~\cite{SuppMat}.

\paragraph*{Experimental realization.}

We demonstrate the SEEQST technique and compare its performance with standard QST on IBM quantum hardware. We conduct experiments on the \textit{ibm\_marrakesh}~\cite{ibm_marrakesh} quantum processor for 2-, 3-, 4-, and 5-qubit states. This device has a two-qubit gate error of $\sim 3.5 \times 10^{-3}$, median relaxation time $T_1 = \qty{200}{\micro\second}$, and median coherence time $T_2 = \qty{115}{\micro\second}$. Since the device does not natively support CNOT gates, we decompose those gates into a controlled-Z (CZ) gate and single-qubit operations. Due to qubit connectivity constraints on the device, extra SWAP operations are required to apply a CZ gate between desired qubits. Each SWAP operation, in turn, introduces three additional CZ gates, further impacting circuit depth and fidelity.

In standard QST, the data is acquired using local unitary operations $\{ I, R_x(\pi/2), R_y(\pi/2)\}^{\otimes N}$ and then collectively post-processed to obtain the entire density matrix. Except for in the one-qubit case, standard QST does not allow for selective reconstruction of density-matrix elements. For the data post-processing in both standard QST and SEEQST, we employ a stochastic gradient descent (SGD) algorithm to minimize the loss function~\cite{Gaikwad2025, ban-pra-1999}
\begin{equation}
    \mathcal{L}(\rho_{\text{ML}} | \bm{d}) = -\sum_{i} d_i \log(p_i) , 
\end{equation}
where $d_i$ denotes measured data from the data set $\bm d$ and $p_i = {\text{Tr}(\Pi_i \rho_{\text{ML}})}$ is the corresponding predicted data with $\Pi_i$ the associated measurement operator and $\rho_{\rm ML}$ the predicted density matrix. To ensure that the GD optimization remains within the space of physical density matrices, we use a Cholesky decomposition parameterization~\cite{Gaikwad2025}: $\rho_{\rm ML} = {T^\dagger T}/{\mathrm{Tr}( T^\dagger T)}$, where $T \in \mathbb{C}^{2^N \times 2^N}$. 

Note that we are optimizing in the full density-matrix space, which is not necessary for SEEQST; we use it here to implement the constraints guaranteeing a physical density matrix. One can easily exponentially reduce the size of the parameter space from $4^N$ to $2^N$ in post-processing using the objective function \textit{min} $\vert \vert A \cdot \bm{\rho}^{S_k} -\bm{d} \vert\vert_{l_2}$, where $A$ is sensing matrix,  $\bm{\rho}^{S_k}$ is a column matrix of the density-matrix elements in the set $S_k$ and $\bm{d}$ is the corresponding data.
In standard QST, the data $\bm d$ is acquired using $3^N$ quantum circuits; in SEEQST, it is obtained from only two quantum circuits per set $S_q$ for estimating selective elements ($2^{N+1} - 1$ circuits for the full $\rho_{\rm ML}$).

\begin{figure}
    \centering 
    \includegraphics[width=1\linewidth]{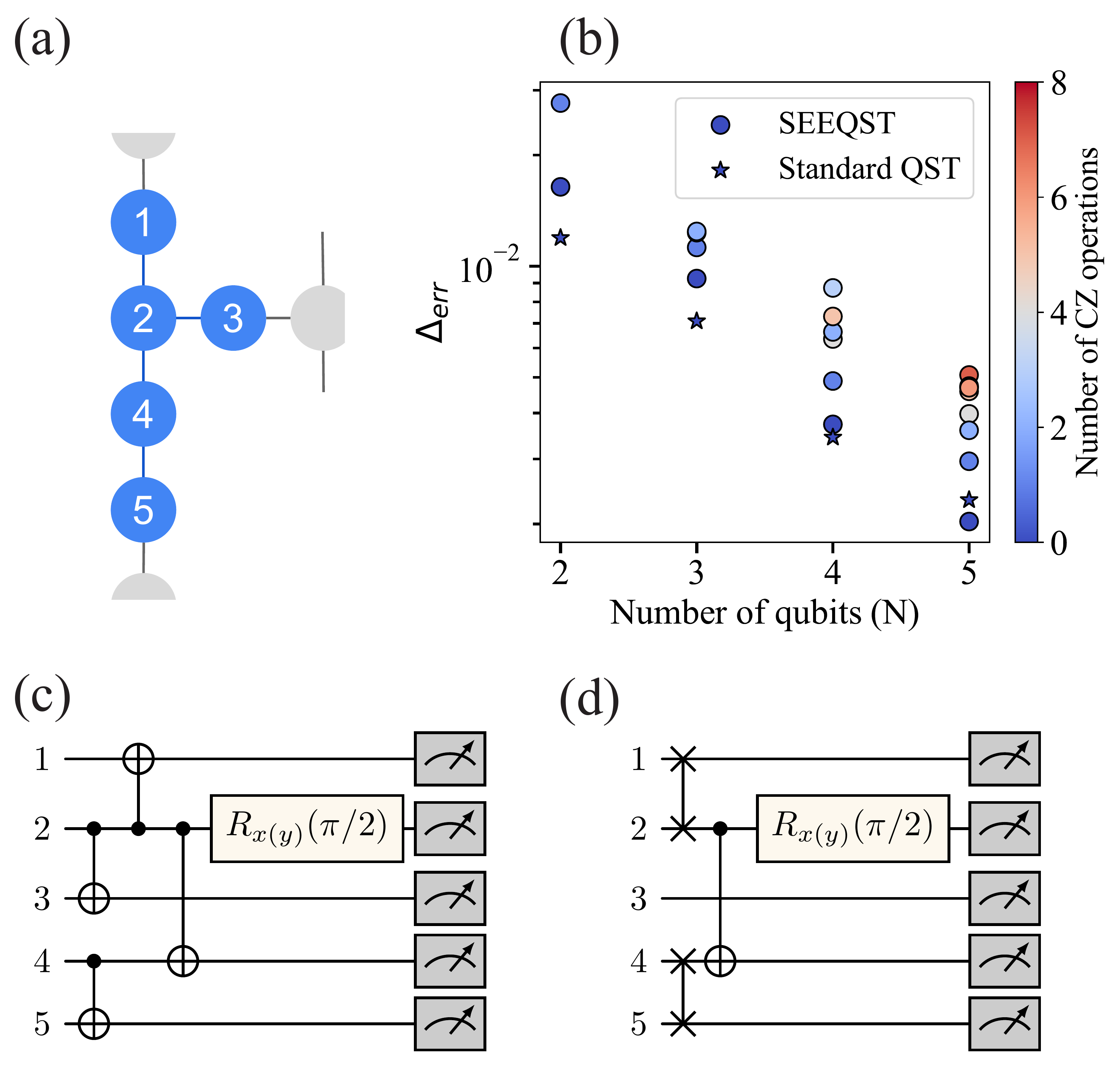} 
    \caption{Experimental implementation.
    (a) Layout of the five qubits used on the \textit{ibm\_marrakesh} quantum processor.
    (b) Average reconstruction error \( \Delta_{\mathrm{err}} \) for the state \(\ket{+y}^{\otimes N}\), as a function of the number of qubits $N$ and the number of CZ operations in the SEEQST circuits. 
    (c) The two circuits for the selective block $S_{32}$ in a 5-qubit state, where all qubits are off-diagonal.
    (d) The two quantum circuits for the selective block $S_{18}$ in a 5-qubit state, where $q_1$ and $q_5$ are off-diagonal. Since they are not directly connected, two additional SWAP operations are required to effectively implement ${\rm CNOT}_{15}$ as required by SEEQST.} 
    \label{fig:experiment} 
\end{figure}

In \figref{fig:experiment}, we present results from applying both standard QST and SEEQST to the \(\ket{+y}^{\otimes N}\) state for systems of up to $N=5$ qubits. We chose this state as a good test case because its density matrix elements are all nonzero, and some of them are complex. In \figpanel{fig:experiment}{b}, we show the average error in estimating an element, considering groups of elements $\cup_{q} S_q$ requiring the same number of CNOT operations in the SEEQST circuits:
\begin{equation}\label{error-definition}
    \Delta_{\mathrm{err}} = \left\langle \left| \rho_k - \rho_k^{\mathrm{ideal}} \right| \right\rangle_{\rho_k \in \cup_{q} S_q}
\end{equation}
For instance, in the two-qubit case (see Fig.~\ref{fig:2qubit}), $\Delta_{err}$ for no CNOT operations is computed over $\cup_{q} S_q = S_1 \cup S_2 \cup S_3 $, while $\Delta_{err}$ for one CNOT operation is calculated using the four elements in the set $S_4$.

We see in \figpanel{fig:experiment}{b} that $\Delta_{\mathrm{err}}$ initially decreases exponentially with the number of qubits, as the size of each matrix element becomes exponentially smaller. However, the error increases with the number of CZ operations required in the SEEQST circuits. This trend breaks down for $N$-qubit GHZ states shown in the SM~\cite{SuppMat}, where the dominant source of error is imperfections in the preparation of such an entangled state. For elements where no CZ gates are involved, the SEEQST circuit resembles standard QST, and the corresponding error closely matches that of standard QST. Thus, the error reflects a trade-off between element size and gate-induced noise.

\begin{table}
\centering
\caption{State fidelity between standard QST, SEEQST, and ideal states for $\left|+y\right\rangle^{\otimes N}$.}
\begin{tabular}{@{}lcccc@{}}
\toprule
QST comparison \: \textbackslash \: $N=$ & 2 & 3 & 4 & 5 \\
\midrule
Standard \& Ideal & 0.992 & 0.988 & 0.980 & 0.974 \\
SEEQST \& Ideal   & 0.985 & 0.977 & 0.970 & 0.952 \\
Standard \& SEEQST & 0.989 & 0.995 & 0.992 & 0.986 \\
\bottomrule
 \label{tbl:plusy}
\end{tabular}
\end{table}


While $\Delta_{err}$ provides an intuitive sense of how circuit complexity affects estimation accuracy, it does not fully capture the impact on overall state reconstruction. Therefore, we compare full state fidelities in Table~\ref{tbl:plusy}. The state fidelity is calculated using as $F(\rho, \sigma) = (\operatorname{tr} \sqrt{\sqrt{\rho} \sigma \sqrt{\rho}})^2$~\cite{jozsa-jmp-1994}. Considering the dominant readout error of approximately $1\%$ per qubit, standard QST itself produces a noisy estimate of the quantum state. The table demonstrates that the reconstructed states from both standard QST and SEEQST are nearly identical, with fidelities ranging between \qty{98.5}{\percent} and \qty{99.5}{\percent}.

Despite the presence of additional gate errors in SEEQST circuits, the method offers a significant advantage over standard QST in runtime. Specifically, SEEQST completes a full 5-qubit QST in just $4.5$ minutes, compared to $17.5$ minutes for standard QST. This illustrates a practical trade-off: for many applications, the marginal increase in error from SEEQST may be acceptable, especially since standard QST already contains inherent noise. Thus, SEEQST can greatly reduce experimental overhead while still yielding high-fidelity state estimates. We further support this in the SM through numerical simulations of SEEQST on full-rank five-qubit states~\cite{SuppMat}.

\paragraph*{Conclusion and outlook.}

We have developed a general method, SEEQST, for selective quantum state tomography of specific sets of $2^N$ elements within an $N$-qubit density matrix. This method requires only two quantum circuits per set, each employing (depending upon the set) at most $N-1$ CNOT gates and one single-qubit rotation. As we have demonstrated both analytically and in experiments with up to $N=5$ qubits, SEEQST enables efficient determination of specific sets of elements without the need for full tomography, and it can perform full QST using only $2^{N+1} - 1$ circuits, fewer than the $3^N$ circuits in standard QST. Furthermore, instead of processing the entire data set at once to infer the quantum state, as standard QST does, SEEQST enables independent computation of subsets, facilitating parallel and simultaneous estimation for even faster full reconstruction. To enable easy implementation, we have made our SEEQST code freely accessible~\cite{SEEQST-python-code}.

A possible extension of SEEQST is integration with threshold-based tomography methods~\cite{binosi-aplquant-2024}. In such methods, one measures the diagonal elements $\rho_{ii}$ and use these to set a significance threshold $t$. This threshold provides a data-driven criterion to reduce the subsequent measurement process: the two SEEQST experiments for a subset $S_k$ are performed only if $\sqrt{\rho_{ii}\rho_{jj}} \ge t$, meaning that at least one of its elements $\rho_{ij}$ is predicted to be non-negligible. 
This reduces experimental overhead without any \textit{a priori} assumptions on the state's structure.


Beyond applying SEEQST to speed up QST in existing systems, future work can also extend the SEEQST framework to enable selective and efficient quantum process tomography (SEEQPT) by leveraging the state-channel duality theorem~\cite{choi-laa-1975, leung-2003, jiang-pra-2013}. However, this straightforward extension would require an equal number of ancilla and system qubits. Performing SEEQPT with a variable number of ancilla qubits---or even without ancilla qubits---poses a non-trivial challenge to be explored in both discrete- and continuous-variable quantum systems. Such protocols would be valuable for efficiently extracting specific information about an unknown quantum process and diagnosing experimental quantum gates to enhance their quality, all without the need for full QPT.


\begin{acknowledgments}

\paragraph*{Acknowledgments.}

We thank the WACQT Quantum Technology Testbed operated by Chalmers Next Labs for help with access to IBM quantum hardware.
We acknowledge support from the Knut and Alice Wallenberg Foundation through the Wallenberg Centre for Quantum Technology (WACQT) and from the EU Flagship on Quantum Technology H2020-FETFLAG-2018-03 project 820363 OpenSuperQ.
AFK is also supported by the Swedish Foundation for Strategic Research (grant numbers FFL21-0279 and FUS21-0063).

\end{acknowledgments}

\bibliography{References}

\onecolumngrid
\clearpage
\setcounter{section}{0}
\renewcommand{\thesection}{S\arabic{section}}
\setcounter{equation}{0}
\renewcommand{\theequation}{S\arabic{equation}}
\setcounter{figure}{0}
\renewcommand{\thefigure}{S\arabic{figure}}
\renewcommand{\thetable}{S\arabic{table}}
\renewcommand{\bibnumfmt}[1]{[S#1]}

\setcounter{page}{1}


\section{Mathematical Induction for Two Experimental Sets}

Here, we show in more detail how the set $\Pi^{S_k}$ can be partitioned into two subsets of mutually commuting observables, denoted by $\mathcal{E}(\Pi^{S_k})$ and $\mathcal{O}(\Pi^{S_k})$.

We follow the convention that the $M$ qubits for which $i_l \neq j_l$ are relabeled as the first $M$ qubits, with the remaining $N-M$ qubits labeled as the last ones. Thereby, only $\sigma_x$ or $\sigma_y$ appear on the first $M$ qubits. The corresponding observables on the first $M$ qubits are denoted by $\mathcal{E}_M$ and $\mathcal{O}_M$, representing $\mathcal{E}(\Pi^{S_k})$ and $\mathcal{O}(\Pi^{S_k})$, respectively. We now prove by induction that: (i) the elements within $\mathcal{E}_M$ commute; (ii) the elements within $\mathcal{O}_M$ commute; (iii) every element of $\mathcal{E}_M$ anti-commutes with every element of $\mathcal{O}_M$.

We start with the simplest case, $M=1$, where
\begin{equation}
\mathcal{E}_1 = \{\sigma_x\}, \quad \mathcal{O}_1 = \{\sigma_y\}.
\end{equation}
Here, (i) and (ii) are trivially fulfilled, and (iii) holds since $\{\sigma_x, \sigma_y\} = 0$.

For $M=2$, we have 
\begin{equation}
\mathcal{E}_2 = \{\sigma_x \otimes \sigma_x,\, \sigma_y \otimes \sigma_y\}, \quad \mathcal{O}_2 = \{\sigma_x \otimes \sigma_y,\, \sigma_y \otimes \sigma_x\}.
\end{equation}
A straightforward calculation shows that elements within $\mathcal{E}_2$ commute, elements within $\mathcal{O}_2$ commute, and any element from $\mathcal{E}_2$ anti-commutes with any element from $\mathcal{O}_2$.

We now assume that for $M-1$ qubits, the following holds:
\begin{align}
[E_i, E_j] &= 0 \quad \quad \quad \text{for all } E_i, E_j \in \mathcal{E}_{M-1}, \notag \\
[O_i, O_j] &= 0 \quad \quad \quad \text{for all } O_i, O_j \in \mathcal{O}_{M-1}, \notag \\
\{E_i, O_j\} &= 0 \quad \quad \quad \text{for all } E_i \in \mathcal{E}_{M-1}, \, O_j \in \mathcal{O}_{M-1}.
\end{align}
We then define
\begin{align}
\mathcal{E}_M &= \left( \mathcal{E}_{M-1} \otimes \sigma_x \right) \cup \left( \mathcal{O}_{M-1} \otimes \sigma_y \right). \\
\mathcal{O}_M &= \left( \mathcal{O}_{M-1} \otimes \sigma_x \right) \cup \left( \mathcal{E}_{M-1} \otimes \sigma_y \right).
\end{align}
Then, for any $E_i, E_j \in \mathcal{E}_{M-1}$, we have
\begin{equation}
[E_i \otimes \sigma_x, E_j \otimes \sigma_x] = [E_i, E_j] \otimes I = 0.
\end{equation}
Similarly, for any $O_i, O_j \in \mathcal{O}_{M-1}$,
\begin{equation}
[O_i \otimes \sigma_y, O_j \otimes \sigma_y] = [O_i, O_j] \otimes I = 0.
\end{equation}
For cross terms, we have
\begin{equation}
[E_i \otimes \sigma_x, O_j \otimes \sigma_y] = E_i O_j \otimes (\sigma_x \sigma_y) - O_j E_i \otimes (\sigma_y \sigma_x) = 0,
\end{equation}
and
\begin{equation}
\{E_i \otimes \sigma_x, O_j \otimes \sigma_x\} = \{E_i, O_j\} \otimes I = 0.
\end{equation}
Thus, the required commutation and anti-commutation properties are preserved going from $M-1$ to $M$, and holds for any $M$ by induction. Finally, extending the result to $N$ qubits by tensoring with $\{\sigma_0, \sigma_z\}^{\otimes (N-M)}$ preserves the commutation relations, since $[\sigma_0, \sigma_z] = 0$. This concludes the proof.

\section{SEEQST of GHZ states}

\begin{figure} 
    \centering 
    \includegraphics[width=0.5\linewidth]{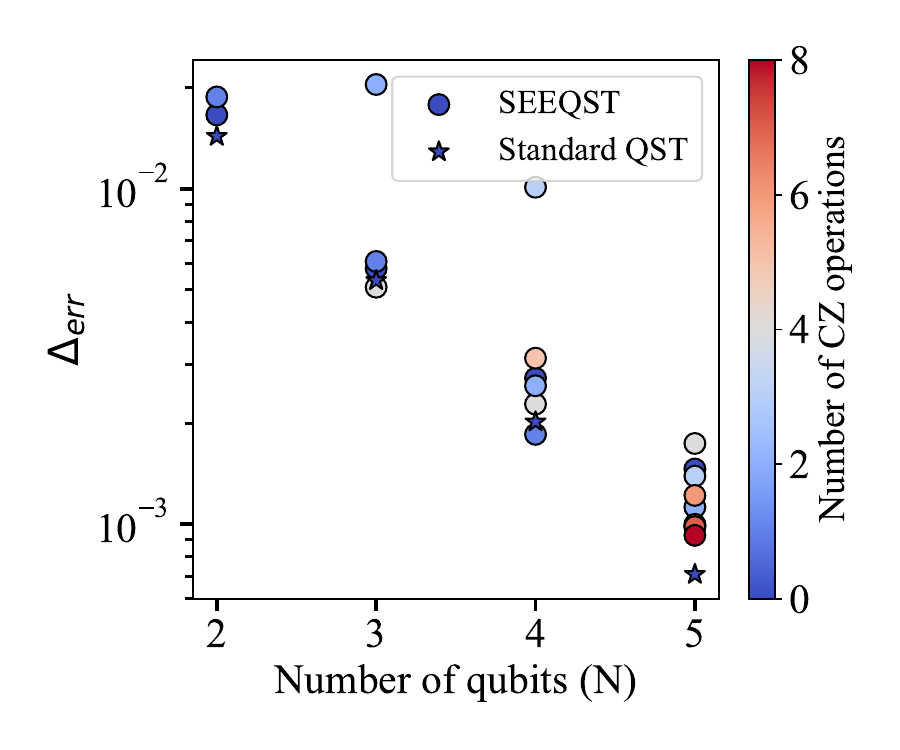} 
    \caption{Average reconstruction error \( \Delta_{\mathrm{err}} \) for $N$-qubit GHZ states on the form $\frac{1}{\sqrt{2}}\mleft(\ket{0}^{\otimes N} - i\ket{1}^{\otimes N}\mright)$, as a function of the number of qubits $N$ and the number of CZ operations in the SEEQST circuits.} 
    \label{fig:experiment2} 
\end{figure}

\begin{table} 
\centering
\caption{State fidelity between standard QST, SEEQST, and ideal states for $\frac{1}{\sqrt{2}}\mleft(\ket{0}^{\otimes N} - i\ket{1}^{\otimes N}\mright)$.}
\begin{tabular}{@{}lrrrr@{}}
\toprule
QST comparison \: \textbackslash \: $N=$ & 2 & 3 & 4 & 5 \\
\midrule
Standard \& Ideal     & 0.977 & 0.973 & 0.966 & 0.956 \\
SEEQST \& Ideal            & 0.983 & 0.972 & 0.966 & 0.953 \\
Standard \& SEEQST     & 0.989 & 0.985 & 0.988 & 0.985 \\
\bottomrule
\label{ghz-fid}
\end{tabular}
\end{table}

In analogy with the experimental results in the main text for the $\ket{+y}^{\otimes N}$ state for systems of up to $N=5$ qubits, we here present in \figref{fig:experiment2} the results of QST and SEEQST applied to an initial GHZ state on the form
\begin{equation}
\ket{\psi_{\text{GHZ}, N}} = \frac{1}{\sqrt{2}}\mleft(\ket{0}^{\otimes N} - i\ket{1}^{\otimes N}\mright).
\end{equation}
The average estimation error, $\Delta_{\mathrm{err}}$, for each density matrix element is calculated using \eqref{error-definition}. As can be seen in \figref{fig:experiment2}, this error decreases exponentially with the number of qubits, consistent with the trend observed in \figref{fig:experiment} for the \(\ket{+y}^{\otimes N}\) state. However, somewhat counterintuitively, the error does not increase with the number of CZ gates for GHZ-like states. This deviation occurs because the main source of error is not the gate operations in the measurement circuits, but the inherent imperfections in preparing the GHZ state, which significantly affect the overall error profile.


In Table~\ref{ghz-fid}, we compare the full state fidelities for the different methods. Due to the dominant readout error of approximately \qty{1}{\percent} per qubit, standard QST already introduces noticeable noise into the reconstructed state. Nonetheless, the results show that SEEQST achieves fidelities comparable to standard QST, with both methods producing nearly indistinguishable states across all tested qubit numbers.

\section*{Circuit Depth Complexity in SEEQST}

In this supplementary section, we analyze the circuit depth complexity in our SEEQST protocol. Specifically, we consider the implementation of $U_{\rm GHZ}^\dagger$ on an arbitrary subset of $M$ qubits within an $N$-qubit system.

Figure~\ref{fig:nqubit} in the main text illustrates a straightforward implementation using a sequential application of $M-1$ CNOT gates. However, various alternative strategies exist for generating GHZ states across different system layouts, often achieving better depth using parallelization of CNOTs. In an all-to-all connectivity scenario, a GHZ state can be created in $\log_2(M)$ depth by leveraging an entanglement-growing strategy: once two qubits are entangled, they can be used to entangle two more, then four, and so on.

For a square lattice, optimal parallelization of CNOTs leads to a circuit depth that scales with $\sqrt{N}$ if all $N$ qubits need to be entangled. For an infinite heavy-hexagonal lattice, the optimal circuit depth $k$ is given by~\cite{KamGHZdepth}
\[
k  = \left\lceil \frac{\sqrt{8N - 3} - 1}{2} \right\rceil.
\]

Applying $U_{\rm GHZ}^\dagger$ directly in these layouts is straightforward for the fully entangled case, but our goal is to generate a GHZ state only within a selected subset of $M$ qubits, which may be arbitrarily positioned within the lattice. This constraint necessitates the use of SWAP operations to bring the target qubits into the required locations. However, SWAP gates introduce additional circuit depth. One way to optimize this is by parallelizing SWAP operations, treating them similarly to controlled-NOT (CNOT) gates. Since finding an optimal SWAP sequence is generally an NP-hard problem, we aim to provide an upper bound on circuit depth.

\begin{figure} 
    \centering 
    \includegraphics[width=0.94 \linewidth]{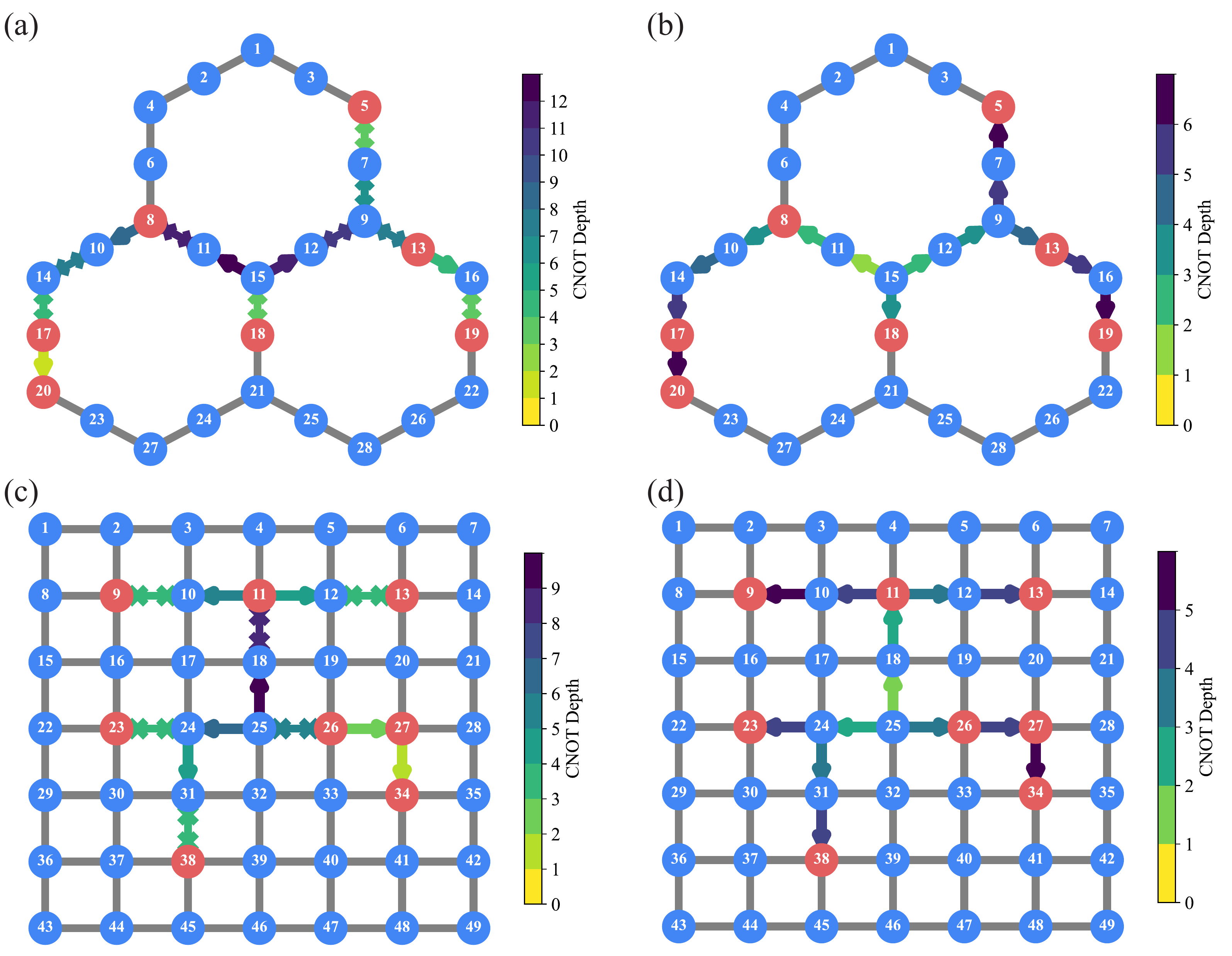} 
    \caption{Circuit depth for SEEQST for different qubit layouts. Red nodes represent off-diagonal qubits. Arrows represent CNOT gates (control $\rightarrow$ target), and SWAP gates are denoted by $\times\!\!-\!\!\times$.
    (a) Implementation on a heavy-hexagon lattice, inspired by (b), which depicts the GHZ-state-generation circuit layout with all red qubits enclosed within boundary $B$.
    (c) Implementation on a square lattice, inspired by the GHZ-state-generation layout shown in (d).} 
    \label{fig:depth} 
\end{figure}

Given an $N$-qubit lattice with $M$ off-diagonal qubits, this subset lies within the $N$-qubit layout and has a boundary $B$. The optimal GHZ generation strategy within this boundary defines the circuit depth $k$. We can use the reverse path of GHZ generation to reach all $M$ qubits. In particular, since a SWAP gate can be decomposed into three native CNOT gates, the worst-case circuit depth is at most $3k$. We provide some examples of these strategies in \figref{fig:depth}, for both heavy-hexagon and square-grid lattices.

Alternative approaches, such as long-range CNOT implementations or GHZ generation via iSWAP gates, could further optimize the depth. Ultimately, users of SEEQST can tailor these methods to their system constraints, as long as they ensure the correct preparation of all eigenstates required by Eqs.~\ref{eq:e_eigenstate} and \ref{eq:o_eigenstate} in the main text.

Note that in our depth calculations, we consider only two-qubit gates, specifically taking CNOT as our baseline metric. We exclude single-qubit gates (including virtual $Z$ gates) since their gate times are signifixantly shorter and their errors are significantly lower than those for two-qubit operations. 

\section*{Experimental data}

In Tables~\ref{tab:datafig3} and \ref{tab:datafig4}, we present the numerical values for the data points plotted in Figs.~\figpanelNoPrefix{fig:experiment}{b} and \ref{fig:experiment2}, respectively.

{
\renewcommand{\arraystretch}{1.2}
\rowcolors{2}{gray!15}{white}
\begin{table}[H]
    \centering
    \rowcolors{2}{gray!15}{white}
    \setlength{\tabcolsep}{0.3pt}
    \renewcommand{\arraystretch}{1.2}
    \begin{tabular}{c c c c c}
        \toprule
        \rowcolor{gray!50}
        Number of CZ operations \textbackslash \: $N=$ & 2 & 3 & 4 & 5 \\
        \midrule
        \multicolumn{5}{c}{\textbf{Standard QST}} \\
        \midrule
        0 & $1.19 \times 10^{-2}$ & $7.09 \times 10^{-3}$ & $3.43 \times 10^{-3}$ & $2.33 \times 10^{-3}$ \\
        \midrule
        \multicolumn{5}{c}{\textbf{SEEQST}} \\
        \midrule
        0 & $1.64 \times 10^{-2}$ & $9.26 \times 10^{-3}$ & $3.73 \times 10^{-3}$ & $2.03 \times 10^{-3}$ \\
        1 & $2.77 \times 10^{-2}$ & $1.12 \times 10^{-2}$ & $6.35 \times 10^{-3}$ & $3.98 \times 10^{-3}$ \\
        2 & ---                  & $1.24 \times 10^{-2}$ & $6.86 \times 10^{-3}$ & $2.96 \times 10^{-3}$ \\
        3 & ---                  & $1.24 \times 10^{-2}$ & $8.73 \times 10^{-3}$ & $3.59 \times 10^{-3}$ \\
        4 & ---                  & ---                  & $4.84 \times 10^{-3}$ & $4.55 \times 10^{-3}$ \\
        5 & ---                  & ---                  & $7.37 \times 10^{-3}$ & $4.70 \times 10^{-3}$ \\
        6 & ---                  & ---                  & ---                  & $5.07 \times 10^{-3}$ \\
        7 & ---                  & ---                  & ---                  & $4.73 \times 10^{-3}$ \\
        8 & ---                  & ---                  & ---                  & $4.68 \times 10^{-3}$ \\
        \bottomrule
    \end{tabular}
    \caption{$\Delta_{\rm err}$ data for the state $\ket{+y}^{\otimes N}$ as illustrated in Fig.~\ref{fig:experiment}(b).}
    \label{tab:datafig3}
\end{table}

\begin{table}[H]
    \centering
    \small
    \setlength{\tabcolsep}{0.3pt} 
    \renewcommand{\arraystretch}{1.3} 
    \begin{tabular}{c c c c c}
        \rowcolor{gray!50}
        \toprule
        Number of CZ operations \textbackslash $N=$ & 2 & 3 & 4 & 5 \\
        \midrule
        \multicolumn{5}{c}{\textbf{Standard QST}} \\
        \midrule
        0 & $1.43 \times 10^{-2}$ & $8.53 \times 10^{-3}$ & $2.02 \times 10^{-3}$ & $7.09 \times 10^{-4}$ \\
        \midrule
        \multicolumn{5}{c}{\textbf{SEEQST}} \\
        \midrule
        0 & $1.67 \times 10^{-2}$ & $7.58 \times 10^{-3}$ & $2.77 \times 10^{-3}$ & $1.46 \times 10^{-3}$ \\
        1 & $1.88 \times 10^{-2}$ & $1.12 \times 10^{-2}$ & $2.83 \times 10^{-3}$ & $8.95 \times 10^{-4}$ \\
        2 & - & $2.04 \times 10^{-2}$ & $2.85 \times 10^{-3}$ & $1.32 \times 10^{-3}$ \\
        3 & - & - & $1.01 \times 10^{-2}$ & $1.39 \times 10^{-3}$ \\
        4 & - & $1.24 \times 10^{-2}$ & $3.13 \times 10^{-3}$ & $2.10 \times 10^{-3}$ \\
        5 & - & - & - & $1.62 \times 10^{-3}$ \\
        6 & - & - & - & $1.21 \times 10^{-3}$ \\
        \bottomrule
    \end{tabular}
    \caption{$\Delta_{\rm err}$ data for the state $\frac{1}{\sqrt{2}}\mleft(\ket{0}^{\otimes N} - i\ket{1}^{\otimes N}\mright)$ as illustrated in Fig.~\ref{fig:experiment2}.}
    \label{tab:datafig4}
\end{table}
}

\section{SEEQST Using Only Single-Qubit Gates}

In \figref{fig:nqubit}, we employed two-qubit gates to measure all observables in $\mathcal{E}(\Pi^{S_k})$ and $\mathcal{O}(\Pi^{S_k})$ in two separate experiments, utilizing the fact that observables within each set commute globally. To implement SEEQST using only single-qubit gates, the relevant observables must commute locally on each qubit. Using \eqref{eq:sqstset}, the circuits required to measure a given $\Pi^{S_k}$ can be constructed as follows:
\begin{equation}\label{eq:singlecircuit}
 \bigotimes_{l=1}^N 
    \begin{cases}
        I, & \text{if } (i_l, j_l) \in \{(0,0), (1,1)\}, \\
        \{ R_x\left(\pi/2\right),\ R_y\left(\pi/2\right) \}, & \text{if } (i_l, j_l) \in \{(0,1), (1,0)\}.
    \end{cases}
\end{equation}

If, for a given set, \( M \) out of \( N \) qubits correspond to odd parity, i.e., $i_k \neq j_k$, then $2^M$ distinct single-qubit rotation circuits are required to fully access the corresponding observables.  
Consequently, the optimal (minimal) total number of distinct circuits across all \( 2^N \) sets is \( 3^N \) in the general case.
Therefore, the SEEQST protocol with only single-qubit gates can be regarded as a constrained variant of standard QST, relying solely on local measurement settings and avoiding entangling operations. This approach has been analyzed in Ref.~\cite{mahesh-arxiv-2025}. 

We analyzed the estimation error \( \Delta_{\mathrm{err}} \) from noiseless simulations of 30 random full-rank density matrices for each \( N = 1 \) to \( 6 \), across varying sample sizes \( S \) (number of measurement settings multiplied by number of shots), for both methods. In all cases, the error exhibited the expected \( 1/\sqrt{S} \) scaling. Additionally, we observed a systematic dependence of the error on the odd-parity index \( M \).

\begin{figure}
    \centering
    \includegraphics[width=0.8\linewidth]{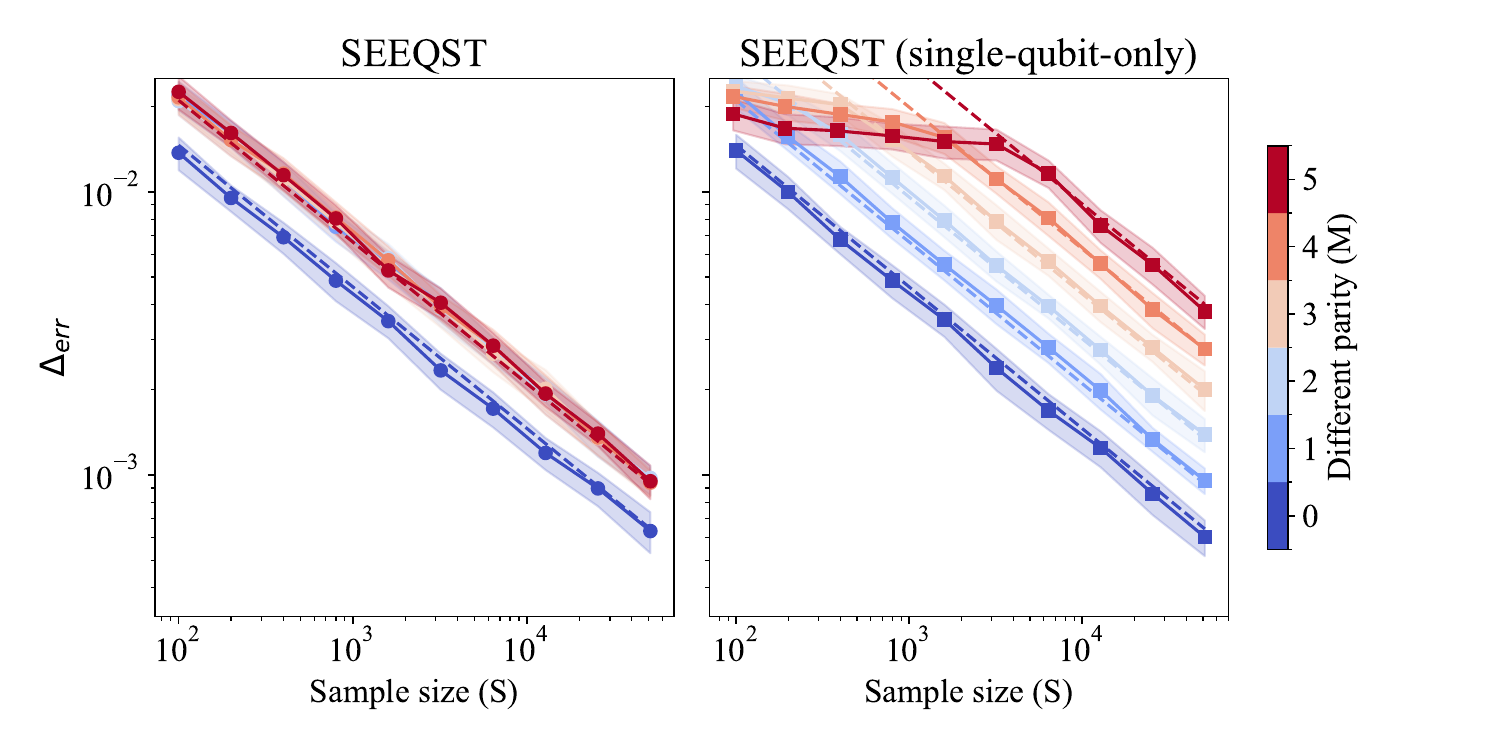}
    \caption{
    Mean estimation error \( \Delta_{\mathrm{err}} \) (with standard deviation shaded) as a function of sample size for the SEEQST protocol on a five-qubit system. The figure demonstrates the statistical advantage of using entangled two-qubit gates compared to implementations only employing single-qubit gates, in terms of the number of copies $S$ of the quantum state \( \rho \) required. For SEEQST with two-qubit gates, estimation performance remains consistent across different block parities \( M \), except for the diagonal block. In contrast, when restricted to single-qubit gates, the estimation error increases with \( M \), indicating degraded scaling. The fitted empirical scaling relation in Eqs.~(\ref{eq:delta_sq})--(\ref{eq:delta_cnot}) (dashed lines) captures the observed behavior within standard deviation, except for with very small sample sizes for higher $M$.
    }
    \label{fig:seeqst_scaling}
\end{figure}

For single-qubit-only SEEQST, the error increases with increasing \( M \), and the error curves appear nearly equally spaced in log scale across different \( M \) values, suggesting exponential dependence on \( M \) as shown in \figref{fig:seeqst_scaling}. 

To capture these features, we adopt the following empirical model for single-qubit-only SEEQST:
\begin{equation}
\Delta_{\mathrm{err}}^{\mathrm{SQ}}(N, M) = \frac{2^{A(N) + B(N) \cdot M}}{\sqrt{S}},
\label{eq:delta_sq}
\end{equation}
where the functions \( A(N) \) and \( B(N) \) capture the system-size dependence. We found that power-law forms for both were sufficient to fit the observed behavior within the standard deviation of the data (except for the case of high $M$ with small sample sizes):
\begin{equation}
A(N) = a_0 + a_1 \cdot N^{a_2}, \quad B(N) = b_1 \cdot N^{b_2},
\label{eq:scaling_functions}
\end{equation}
with fitted coefficients
\[
a_0 = -0.9177, \quad a_1 = -0.24734, \quad a_2 = 1.2529, \quad b_1 = 0.6358, \quad b_2 = -0.1168.
\]

In contrast, our SEEQST protocol employing entangling CNOT gates shows no such dependence on \( M \geq 1 \). Its estimation error is empirically related to the single-qubit model as
\begin{equation}
\Delta_{\mathrm{err}}^{\mathrm{CNOT}}(N, M) =
\begin{cases}
\Delta_{\mathrm{err}}^{\mathrm{SQ}}(N, M = 1), & \text{for } M \geq 1, \\
\Delta_{\mathrm{err}}^{\mathrm{SQ}}(N, M = 0), & \text{for } M = 0.
\end{cases}
\label{eq:delta_cnot}
\end{equation}

This highlights the statistical advantage of utilizing entangling gates in SEEQST.

\section{Numerical simulation}

\begin{figure}
    \centering
    \includegraphics[width=\linewidth]{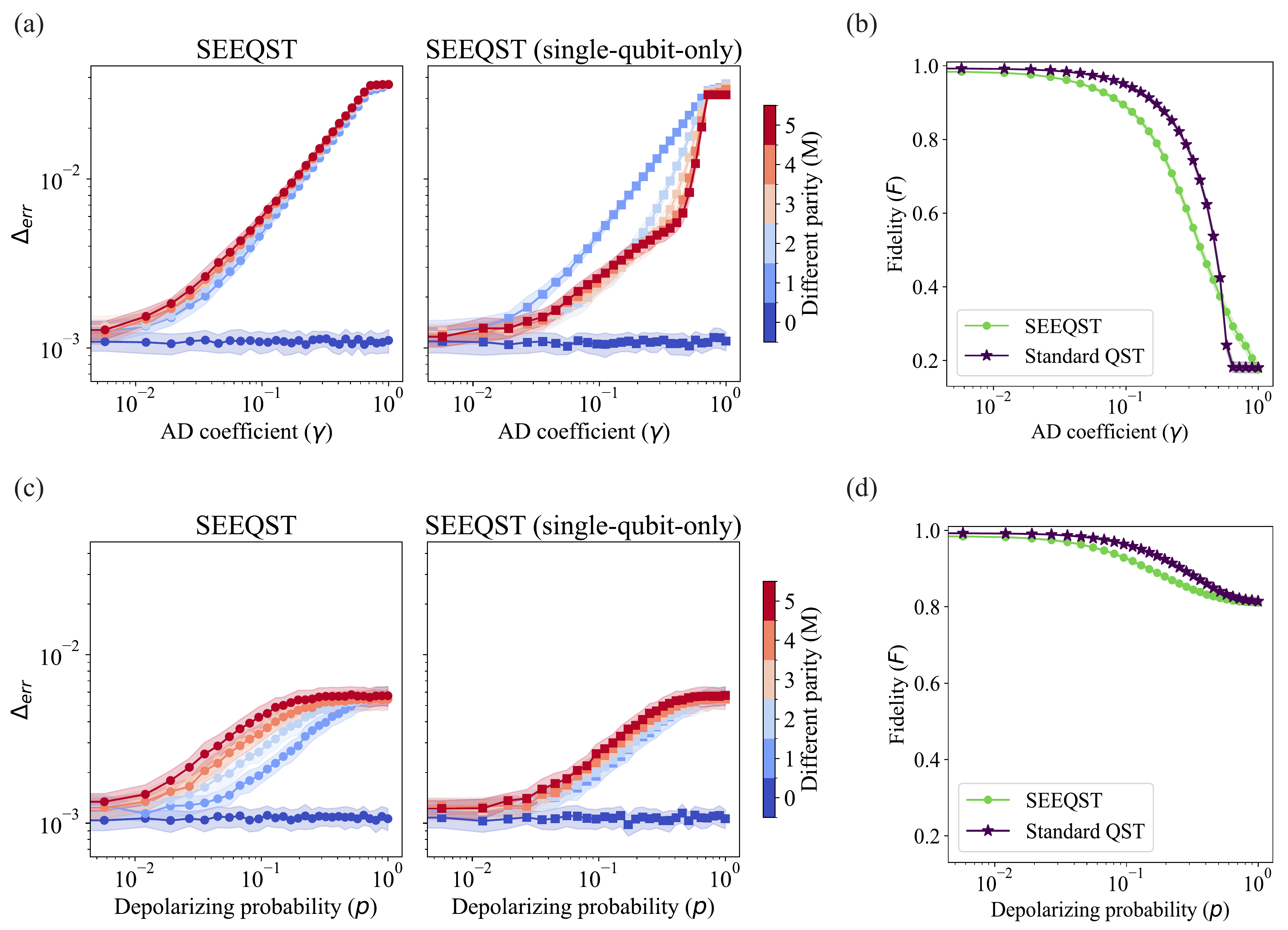}
    \caption{Numerical simulation of SEEQST performance under noise. The effect of (a) amplitude damping and (c) depolarizing noise on the estimation error $\Delta_{\rm err}$ for different SEEQST subsets, categorized by the different parity $M$. Left panels show results for the standard SEEQST protocol, while right panels show the performance for its single-qubit gates only variant. 
    (b, d) Impact of amplitude damping and depolarizing noise on the state fidelity $F$, comparing SEEQST with standard QST. All simulations are for full-rank five-qubit states and 16,384 shots per measurement setting. Solid lines represent the mean over multiple simulation runs, and the shaded regions show the standard deviation.}
    \label{fig:noise}
\end{figure}

To systematically evaluate the performance of SEEQST in noisy environments, we conducted numerical simulations on full-rank five-qubit states, using two standard noise models: amplitude damping and depolarizing noise. In each case, noise was introduced by applying errors with a given probability or coefficient after every layer of parallel gates in the measurement circuits. All results were obtained using 16,384 shots per experiment, with state reconstruction performed via a gradient-descent algorithm, as described in the main text. The results are summarized in \figref{fig:noise}.

We see from the left panels in \figpanel{fig:noise}{a,c} that, in the case of standard SEEQST, the estimation error $\Delta_{\rm err}$ depends more strongly on the parity parameter $M$ under depolarizing noise than in the presence of amplitude damping. From the right panels in \figpanel{fig:noise}{a,c}, we further observe that the opposite is the case for $\Delta_{\rm err}$ in single-qubit-only SEEQST; here, the dependence on $M$ is stronger with amplitude damping than with depolarizing noise. Overall, the standard SEEQST protocol is more noise-sensitive than its single-qubit-only variant, highlighting that while entangling circuits offer a statistical advantage, they are also more susceptible to noise. In terms of overall performance, the state fidelity of SEEQST is slightly lower than for standard QST under both amplitude damping and depolarizing noise, as shown in \figpanel{fig:noise}{b,d}.

\begin{figure} 
    \centering 
    \includegraphics[width=0.45\linewidth]{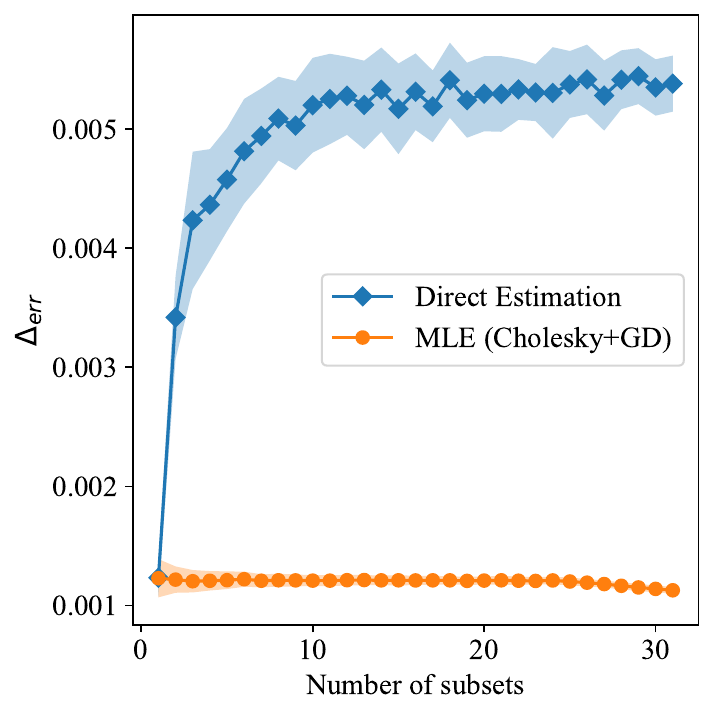} 
    \caption{\(\Delta_{\rm err}\) as a function of the number of non-diagonal Pauli subsets involved in the reconstruction, averaged over 30 randomly generated full-rank 5-qubit states using simulated (16,384 shots per setting). Maximum likelihood estimation with Cholesky decomposition and gradient descent (GD) maintains a stable error across subset counts due to CPTP constraints, whereas direct estimation shows increasing \(\Delta_{\rm err}\) with more subsets.} 
    \label{fig:subset_vs_error}
\end{figure}

As described in the main text, we employ two approaches for reconstructing the density matrix: (i) maximum likelihood estimation (MLE) with Cholesky decomposition and gradient-descent optimization over the full parameter space of the density matrix, and (ii) direct estimation, where expectation values \(\langle P_i \rangle\) of Pauli observables are estimated using measured population frequencies as probabilities for the relevant subset of observables. While direct estimation is significantly more computationally efficient, taking on average \(10^{-4}\) seconds per state on a standard quadcore CPU compared to \(0.3\) seconds for MLE for five-qubit states, it exhibits critical limitations. Figure~\ref{fig:subset_vs_error} illustrates this drawback.

We compute \(\Delta_{\rm err}\) across 30 randomly sampled full-rank quantum states for varying numbers of randomly selected subsets (excluding the diagonal subset due to its trivial nature and distinct sample-size scaling). For each number of subsets involved, results are averaged over 10 random selections. The plot shows that while the error for MLE remains nearly constant with increasing subsets due to the enforcement of complete positivity and trace preservation (CPTP), direct estimation matches MLE performance only for a single subset, and deteriorates quickly as more subsets are included. Although this analysis uses simulated data with \(16{,}384\) shots, the effect is expected to be stronger in experiments, where additional noise sources exist. Indeed, as we observed in our IBM hardware experiments,direct estimation exhibited deviations from MLE even when applied to a single subset.

\end{document}